\documentclass[10pt,journal,compsoc]{IEEEtran}
\ifCLASSOPTIONcompsoc
  \usepackage[nocompress]{cite}
\else
  \usepackage{cite}
\fi
\ifCLASSINFOpdf
  \usepackage[pdftex]{graphicx}
  \graphicspath{{../images/}}
  \DeclareGraphicsExtensions{.pdf,.jpeg,.png}
\else
\fi
\usepackage{amsmath}
\usepackage{algorithmic}

\usepackage{array}

\ifCLASSOPTIONcompsoc
 \usepackage[caption=false,font=footnotesize,labelfont=sf,textfont=sf]{subfig}
\else
 \usepackage[caption=false,font=footnotesize]{subfig}
\fi
\usepackage{url}

\usepackage[table]{xcolor}
\usepackage{booktabs}
\usepackage{multirow}
\usepackage{multicol}
\usepackage{siunitx}
\usepackage{comment}
\usepackage[acronym, nopostdot]{glossaries}
\glsdisablehyper
\usepackage{soul}
\usepackage{eso-pic}
\newcommand{\myCommentDMQ}[1]{}
\newcommand{\myCommentABG}[1]{}
\newcommand{\revisedChanges}[1]{#1}
\newcommand{\revisedChangesTwo}[1]{#1}
\newcommand{\removedChanges}[1]{}
\newcommand{\removedChangesTwo}[1]{}

\newcommand{\instbit}[1]{\mbox{\scriptsize #1}}
\newcommand{\instbitrange}[2]{~\instbit{#1} \hfill \instbit{#2}~}

\newcommand{\CASE}[1]{\STATE \textbf{case} #1\textbf{:} \begin{ALC@g}}
\newcommand{\CASECOMMENT}[2]{\STATE \textbf{case} #1\textbf{:} \COMMENT{#2}\begin{ALC@g}}
\newcommand{\ENDCASE}{\end{ALC@g}}

\newcommand{\DEFAULT}{\STATE \textbf{default:} \begin{ALC@g}}
\newcommand{\DEFAULTCOMMENT}[1]{\STATE \textbf{default:} \COMMENT{#1}\begin{ALC@g}}
\newcommand{\ENDDEFAULT}{\end{ALC@g}}
\newcommand{\DEFAULTLINE}[1]{\STATE \textbf{default:} }

\begin{document}


\newacronym{ALU}{ALU}{Arithmetic Logic Unit}
\newacronym{CNN}{CNN}{Convolutional Neural Network}
\newacronym{DNN}{DNN}{Deep Neural Network}
\newacronym{DSP}{DSP}{Digital Signal Processing}
\newacronym{FF}{FF}{Flip-flop}
\newacronym{FPGA}{FPGA}{Field-Programmable Gate Array}
\newacronym{ASIC}{ASIC}{Application-Specific Integrated Circuit}
\newacronym{FPU}{FPU}{Floating-Point Unit}
\newacronym{GEMM}{GEMM}{General Matrix Multiplication}
\newacronym{ISA}{ISA}{Instruction Set Architecture}
\newacronym{LUT}{LUT}{Lookup Table}
\newacronym{MAC}{MAC}{Multiply-Accumulate}
\newacronym{MaxAbsE}{MaxAbsE}{Maximum Absolute Error}
\newacronym{MSE}{MSE}{Mean Squared Error}
\newacronym{NaN}{NaN}{Not a Number}
\newacronym{NaR}{NaR}{Not-a-Real}
\newacronym{NPB}{NPB}{NAS Parallel Benchmark}
\newacronym{PAU}{PAU}{Posit Arithmetic Unit}
\newacronym{RISC}{RISC}{Reduced Instruction Set Computer}
\newacronym{RTL}{RTL}{Register-Transfer Level}
\newacronym{SoC}{SoC}{System on a Chip}
\newacronym{HPC}{HPC}{High-Performance Computing}

%
\title{Big-PERCIVAL: Exploring the Native Use of 64-Bit Posit Arithmetic in Scientific Computing}
%
%
%
%

\author{David~Mallasén,
        Alberto~A.~Del~Barrio,~\IEEEmembership{Senior~Member,~IEEE}
        and~Manuel~Prieto-Matias
\thanks{All authors are with the Facultad de Informática, Universidad Complutense de Madrid, 28040 Madrid, Spain.\protect\\
E-mails: \{dmallase, abarriog, mpmatias\}@ucm.es}
\thanks{Manuscript received -; revised -.}}

\AddToShipoutPictureBG*{%
  \AtPageUpperLeft{%
    \setlength\unitlength{1in}%
    \hspace*{\dimexpr0.5\paperwidth\relax}
    \makebox(0,-0.5)[c]{\footnotesize This article has been accepted for publication in a future issue of this journal, but has not been fully edited. Content may change prior to final publication.}%
}}
\AddToShipoutPictureBG*{%
  \AtPageUpperLeft{%
    \setlength\unitlength{1in}%
    \hspace*{\dimexpr0.5\paperwidth\relax}
    \makebox(0,-0.8)[c]{\footnotesize Citation information: DOI  10.1109/TC.2024.3377890, IEEE Transactions on Computers}%
}}
\AddToShipoutPictureBG*{%
  \AtPageLowerLeft{%
    \setlength\unitlength{1in}%
    \hspace*{\dimexpr0.5\paperwidth\relax}
    \makebox(0,0.8)[c]{\footnotesize This work is licensed under a Creative Commons Attribution-NonCommercial-NoDerivatives 4.0 License.}%
}}
\AddToShipoutPictureBG*{%
  \AtPageLowerLeft{%
    \setlength\unitlength{1in}%
    \hspace*{\dimexpr0.5\paperwidth\relax}
    \makebox(0,0.5)[c]{\footnotesize For more information, see https://creativecommons.org/licenses/by-nc-nd/4.0/}%
}}

\IEEEtitleabstractindextext{%
\begin{abstract}
The accuracy requirements in many scientific computing workloads result in the use of double-precision floating-point arithmetic in the execution kernels. Nevertheless, emerging real-number representations, such as posit arithmetic, show promise in delivering even higher accuracy in such computations. In this work, we explore the native use of 64-bit posits in a series of numerical benchmarks \removedChanges{extracted from the PolyBench collection} and compare their timing performance, accuracy and hardware cost to IEEE 754 doubles. \revisedChanges{In addition, we also study the conjugate gradient method for numerically solving systems of linear equations in real-world applications.} For this, we extend the PERCIVAL RISC\nobreakdash-V core and the Xposit custom RISC\nobreakdash-V extension with posit64 and quire operations. Results show that posit64 can\removedChanges{ execute as fast as doubles, while also} obtain up to 4 orders of magnitude lower mean square error \removedChanges{and up to 3 orders of magnitude lower maximum absolute error} \revisedChanges{than doubles. This leads to a reduction in the number of iterations required for convergence in some iterative solvers.} However, leveraging the quire accumulator register can limit the order of some operations such as matrix multiplications. Furthermore, detailed FPGA \revisedChangesTwo{and ASIC} synthesis results highlight the significant hardware cost of 64-bit posit arithmetic and quire. Despite this, the large accuracy improvements achieved with the same memory bandwidth suggest that posit arithmetic may provide a potential alternative representation for scientific computing.
\end{abstract}

\begin{IEEEkeywords}
Arithmetic, Posit, IEEE-754, Floating point, Scientific computing, RISC\nobreakdash-V, CPU, Matrix multiplication, PolyBench.
\end{IEEEkeywords}}

\maketitle

\IEEEdisplaynontitleabstractindextext

%
\IEEEpeerreviewmaketitle

\IEEEraisesectionheading{\section{Introduction}\label{sec:introduction}}

%
%
%
%

\IEEEPARstart{R}{eal-number} arithmetic is at the core of many scientific workloads. Physical constants, data from sensors, and in general most inputs to experimental applications have to be represented accurately in a computer. Moreover, this accuracy has to be maintained throughout the execution of the algorithms that are at the root of scientific computing. Even minor errors can result in significant consequences, potentially leading to incorrect predictions of the behavior of a system or inaccurate solutions to differential equations and optimization problems.

The most widely used representation of real numbers in a computer is the IEEE 754 standard for floating-point arithmetic~\cite{ieeecomputersociety2019IEEE}. Although this standard is considered to be a robust and reliable method for representing and operating with real numbers on a computer, it is not perfect. For instance, its results can be inconsistent across platforms, it does not ensure the associative property of additions and multiplications, it has signed zeros, and there is an excess of \gls{NaN} representations.

In the past years, other alternatives to this floating-point format have emerged. Some of these new arithmetic representations have been implemented by large technological companies, especially in the machine learning domain. Examples of this are Google’s bfloat16~\cite{BFloat16} or Nvidia’s TensorFloat~\cite{kharya2020NVIDIA}. In scientific computing, the solution to the accuracy requirements is to use wider floating-point representations such as double-precision floats. However, another solution is to explore emerging floating-point representations that provide more accuracy bits. One of the most promising alternatives for this purpose is posit numbers, which we study in this work.

Targeting this goal, we have extended the PERCIVAL posit RISC\nobreakdash-V core~\cite{mallasen2022PERCIVAL} to support 64-bit posits, as well as including a more diverse and flexible design. This has allowed us to explore the native use of this arithmetic \revisedChanges{with a larger bit-width} both at the hardware-cost level and at the accuracy and performance levels. For the first part, we have performed \gls{FPGA} \revisedChangesTwo{and \gls{ASIC}} synthesis of different configurations of Big-PERCIVAL, and given a detailed analysis of the results. For the accuracy and performance comparison of the arithmetics, we have added posit support for the PolyBench benchmark suite~\cite{pouchet2016PolyBench}. \revisedChanges{In addition, we have also studied the accuracy of iterative linear solvers (conjugate gradient and biconjugate gradient) as well-known examples of widely used applications from science and engineering that can benefit from higher accuracy}~\cite{durand2022Accelerating}. Results were measured on Big-PERCIVAL running on the Genesys II \gls{FPGA} board.

Our main contributions can be summarized in the following:
\begin{itemize}
    \item We present Big-PERCIVAL, an extension of the PERCIVAL\footnote{\url{https://github.com/artecs-group/PERCIVAL}}~\cite{mallasen2022PERCIVAL} posit RISC\nobreakdash-V core which adds posit64 operations and increased flexibility. In particular, we support standard posit addition, subtraction, multiplication, division and square root, conversions to and from integer numbers, comparison operations and register move instructions. Optionally we also support quire operations and logarithmic-approximate multiplication, division and square root units.
    
    \item Detailed \gls{FPGA} \revisedChangesTwo{and \gls{ASIC}} synthesis results of the \gls{PAU} in Big-PERCIVAL showcase the area impact of posit arithmetic and quire in different configurations of the core. These results are compared with the FPNew IEEE 754 \gls{FPU}~\cite{mach2021FPnew}. An analysis of the individual units in the \gls{PAU} gives insight into how the hardware resources are distributed among the different operations.
    
    \item Compiler support for posit64 numbers in the Xposit custom RISC\nobreakdash-V extension in LLVM\footnote{\url{https://github.com/artecs-group/llvm-xposit}}. This allows for easily embedding posit and quire instructions, including loads and stores, into C code.

    \item PolyBench benchmark results provide insight into how \revisedChanges{posit32 and} posit64 numbers compare to IEEE 754 \revisedChanges{floats and} doubles in terms of timing performance and accuracy. In particular, the impact of the quire accumulator register is also studied. Results show that 64-bit posits can provide up to 4 orders of magnitude lower \gls{MSE} and up to 3 orders of magnitude lower \gls{MaxAbsE} than 64-bit doubles.\removedChanges{ Furthermore, we show that there is no performance penalty when executing these benchmarks.} This provides improved accuracy with the same \removedChanges{ performance and }memory bandwidth as doubles.

    \item \revisedChanges{Iterative linear equation solvers, namely the conjugate gradient and biconjugate gradient algorithms, showcase how posit64 can reduce the number of iterations needed to reach a certain tolerance margin when solving large ill-conditioned systems which are frequent in scientific-computing and engineering problems.}

    \item Detailed analysis on how leveraging the quire affects the order in which some operations are executed. For example, the execution of the \gls{GEMM} kernel using a dot-product or memory-aware method impacts the final timing performance and accuracy.
\end{itemize}

The rest of the paper is organized as follows: Section~\ref{sec:posit_arithmetic} introduces posit arithmetic. Related works on RISC\nobreakdash-V cores, the use of posits in \gls{HPC} and theoretical studies on posit64 are presented in Section~\ref{sec:related_work}. In Section~\ref{sec:posit_core} we describe the novelties on Big-PERCIVAL and the Xposit custom RISC\nobreakdash-V extension. The \removedChangesTwo{FPGA }synthesis results of the core and the individual posit64 units are analyzed in Section~\ref{sec:synthesis_results}. Benchmark results using PolyBench targeting accuracy and timing performance are shown in Section~\ref{sec:polybench}, followed by \revisedChanges{the conjugate gradient use-case} in Section~\ref{sec:cg}, and a more exhaustive analysis of the GEMM kernel in Section~\ref{sec:gemm}. Finally, Section~\ref{sec:conclusions} concludes this work.

\section{Posit Arithmetic} \label{sec:posit_arithmetic}

The posit number standard~\cite{positworkinggroup2022Standard}, defines a posit configuration from its total bit-width $n$. This allows for any posit sizes, but in the literature, the most common ones are the byte-aligned posit8, posit16, posit32, and posit64 configurations.

\begin{figure}[htbp]
    \centering
    \includegraphics[width=\columnwidth]{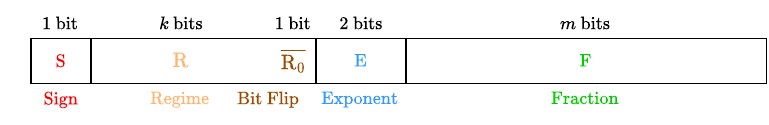}
    \caption{Posit format with sign, regime, exponent, and fraction fields.}
    \label{fig:posit_format}
\end{figure}

One of the main benefits of posit arithmetic is that it does not have a variety of special cases that have to be checked. Posits have only two special cases. The value zero is represented as $\texttt{0}\cdots\texttt{0}$, and the \gls{NaR} is represented as $\texttt{10}\cdots\texttt{0}$. The rest of the bit patterns are composed of the four fields shown in Figure~\ref{fig:posit_format}.

These four bit-fields are:
\begin{itemize}
    \item The sign bit S, the value of which is $s=0$ if the value is positive or $s=1$ if the value is negative.
    \item The variable-length regime field R, which consists of a series of $k$ bits equal to $R_0$ and terminated either by $1-R_0$ or the end of the posit. This field represents a long-range scaling factor $r$ given by:
    \begin{equation*}
        r = \left\{
    	\begin{array}{ll}
    		-k & \mbox{if } R_0 = 0 \\
    		k-1 & \mbox{if } R_0 = 1
    	\end{array}
    	\right.
    \end{equation*}
    \item The exponent field $E$, consisting of at most $2$ bits. This field encodes an integer unbiased value $e$. Since the regime field is variable-length, one or both of the exponent bits may be located after the least significant bit of the posit. In this case, those bits will have the value $0$.
    \item The variable-length fraction field F, which is formed by the $m$ remaining bits. Its value $f$ will be given by dividing the unsigned integer $F$ by $2^m$ and therefore $0 \leq f < 1$.
\end{itemize}

From these fields, we can calculate the real value $p$ of a generic posit as:
\begin{equation} \label{eq:posit_value}
    p = ((1 - 3s) + f)\times 2^{(1-2s)\times(4r + e + s)}.
\end{equation}
This is the most efficient decoding of posits, as shown by~\cite{murillo2022Comparing,uguen2019Evaluating}. The most notable differences in this value representation between posit arithmetic and the IEEE 754 floating-point standard are the existence of the variable-length regime, the use of an unbiased exponent, and the value of the hidden bits~\cite{murillo2022Comparing}. In floating-point arithmetic, the hidden bit is fixed to $1$, except for the subnormal numbers, when it is fixed to $0$. However, in posit arithmetic, it is kept as $1$ if the number is positive, or changed to $-2$ if the number is negative.

\begin{figure}[htbp]
    \centering
    \includegraphics[width=\columnwidth]{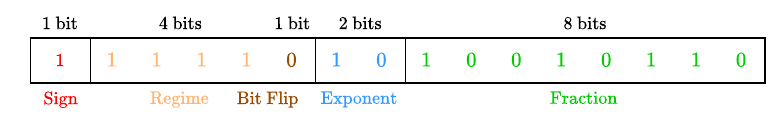}
    \caption{Decoding example of a posit16.}
    \label{fig:posit_format_example}
\end{figure}

As an example, let \texttt{1111101010010110} be the binary encoding of a Posit16 (Figure~\ref{fig:posit_format_example}). The first bit $s=\texttt{1}$ indicates a negative number. The regime field \texttt{11110} gives $k=4$ and therefore $r=3$. The next two bits \texttt{10} represent the exponent $e=2$. Finally, the remaining $m=8$ bits, \texttt{10010110}, encode a fraction value of $f=150/2^8=0.5859375$. Hence, from Equation~(\ref{eq:posit_value}) we conclude that $\texttt{1111101010010110}\equiv (-2 + 0.5859375)\times 2^{-(4\cdot3+2+1)} = -0.000043154$.

The variable-length regime field acts as a long-range dynamic exponent, as can be seen in Equation~(\ref{eq:posit_value}), where it is multiplied by 4 or, equivalently, shifted left by the two exponent bits. Since the regime and the fraction are dynamic fields, they allow for more flexibility in the trade-off between accuracy and dynamic range that can be achieved by a posit. If the regime field occupies more bits, it represents larger numbers at the cost of lower accuracy. On the other hand, when the regime field consists of fewer bits, posits have higher accuracy in the neighborhoods of $\pm 1$.

In posit arithmetic, \gls{NaR} has a unique representation that maps to the most negative 2's complement signed integer. Consequently, if used in comparison operations, it results in less than all other posits and equal to itself. Moreover, the rest of the posit values follow the same ordering as their corresponding bit representations. These characteristics allow posit numbers to be compared as if they were 2's complement signed integers, eliminating additional hardware for posit comparison operations.

Posit arithmetic also includes fused operations using the quire, a $16n$-bit fixed-point 2's complement register. This special accumulation register allows for the execution of up to $2^{31} - 1$ \gls{MAC} operations without intermediate rounding or accuracy loss. These operations are very common when computing dot products, matrix multiplications, or other more complex algorithms. The additional accuracy that can be achieved using the quire can allow the execution of these algorithms with narrower posit configurations~\cite{chaurasiya2018Parameterized,klower2019Posits,neves2020Dynamic}, thus avoiding the limits that can occur in memory bandwidth.

Currently, one of the main drawbacks of posit arithmetic is its higher area cost~\cite{mallasen2022PERCIVAL}. For an accurate comparison between posits and floats, the \gls{FPU} must be IEEE 754 compliant instead of being limited to normal floats only. Authors in~\cite{guntoro2020Next} state that posit hardware is slightly more expensive than floating-point hardware that does not take into account subnormal numbers. Moreover, adding a wide quire accumulator register further increases the area cost of implementing these fused operations.

\section{Related Work} \label{sec:related_work}

Common open-source application-class RISC\nobreakdash-V cores include support for the F and D RISC\nobreakdash-V extensions for IEEE-754 single- and double-precision floating-point numbers (which are part of the G compilation). Some of the most notable ones include the Rocket~\cite{asanovic2016Rocket}, the CVA6 (ex. Ariane)~\cite{zaruba2019Cost}, the Berkeley Out-of-Order Machine (BOOM)~\cite{celio2015Berkeley} or the SHAKTI C-Class processor~\cite{gala2016SHAKTI}.

There have been several previous proposals of including some levels of posit arithmetic capabilities into RISC\nobreakdash-V cores. PERC~\cite{arunkumar2020PERC} and PERI~\cite{tiwari2021PERI} included \glspl{PAU} into the Rocket core and the SHAKTI C-class core, respectively. These proposals were constrained by the F and D extensions and thus did not include quire support. CLARINET~\cite{sharma2022CLARINET} added fused \gls{MAC}, and fused divide and accumulate operators using the quire to an RV32IMAFC RISC\nobreakdash-V core.

In~\cite{cococcioni2021Lightweight} authors store IEEE floats in memory by first converting them to the posit representation. This allows storage of the real-number values with a lower bit-width while performing the computations using the IEEE 754 \gls{FPU}. More recently a posit dot-product unit was presented in~\cite{li2023PDPU}. This open-source implementation allows to perform high-throughput dot-products for deep learning applications. 

In the \gls{HPC} field, \cite{klower2019Posits} explores the use of 16-bit posits in a shallow water model as an example of a medium-complexity climate application. In~\cite{chien2020Posit} authors adapt the \gls{NPB} to 32-bit posits, concluding that they obtain between 0.6 and 1.4 additional decimal digits of accuracy. However, the software emulation resulted in a $4\times$ to $19\times$ performance overhead. An evaluation of the numerical stability of posits for solving linear systems is presented in~\cite{buoncristiani2020Evaluating}. Here, the authors find that there is no big difference between posits and floats in the native range of the matrices that they test. However, when re-scaling the matrices to optimize the use of the posit representation, they obtained 4 extra bits of precision with 32-bit numbers and 2 extra bits with 16-bit numbers. The use of 32-bit posits for the conjugate gradient method is also studied in~\cite{mallasenquintana2022Leveraging}. A \gls{PAU} called POSAR is presented in~\cite{ciocirlan2021Accuracy}. This work analyzes 32-bit posits on one \gls{NPB} scientific application and a \gls{CNN} inference model.

The use of 64-bit posits is studied theoretically in~\cite{dedinechin2019Posits} and in the initial presentation of this arithmetic~\cite{gustafson2017Beating}. At the hardware level, synthesis results of some posit64 operators are given in~\cite{forget2021Comparing, jean2021PFMA, uguen2019Evaluating}, and a \gls{GEMM} accelerator which also supports 64-bit posits is presented in~\cite{ledoux2022Generator}. \revisedChanges{A more complex architecture with a reconfigurable posit tensor unit supporting posit64 is introduced }in~\cite{neves2021Reconfigurable}.

\section{Big-PERCIVAL Core} \label{sec:posit_core}

The PERCIVAL~\cite{mallasen2022PERCIVAL} posit RISC\nobreakdash-V core is based on the application-level CVA6 core\cite{zaruba2019Cost}, to which it adds tightly coupled 32-bit posit and quire capabilities. This allows for both posits and IEEE 754 floats to coexist natively in hardware, which is essential for the evaluation of their strengths and weaknesses.

\begin{figure*}
    \centering
    \includegraphics[width=0.9\textwidth]{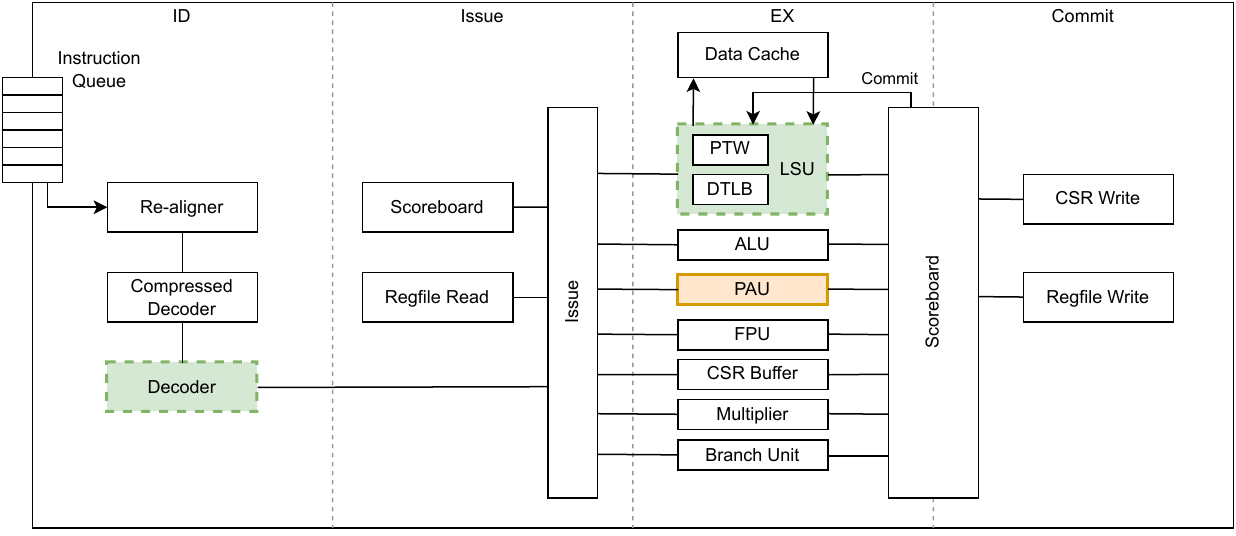}
    \caption{Block diagram of Big-PERCIVAL. Extended modules are highlighted in green. The \gls{PAU} with the new posit64 units is shown in orange.}
    \label{fig:Big-PERCIVAL}
\end{figure*}

The Big-PERCIVAL core we present in this paper is a modified version of PERCIVAL that supports either posit64 numbers or posit32 (see Figure~\ref{fig:Big-PERCIVAL}), although in this work we have focused on the 64-bit version. This includes a parameterized $16n$-bit quire, which results in 1024 bits for posit64 numbers. We also provide a 32-entry posit register file for either posit32 or posit64. The decision between single- or double-precision posits is taken at synthesis time.

\begin{figure*}
    \centering
    \includegraphics[width=0.9\textwidth]{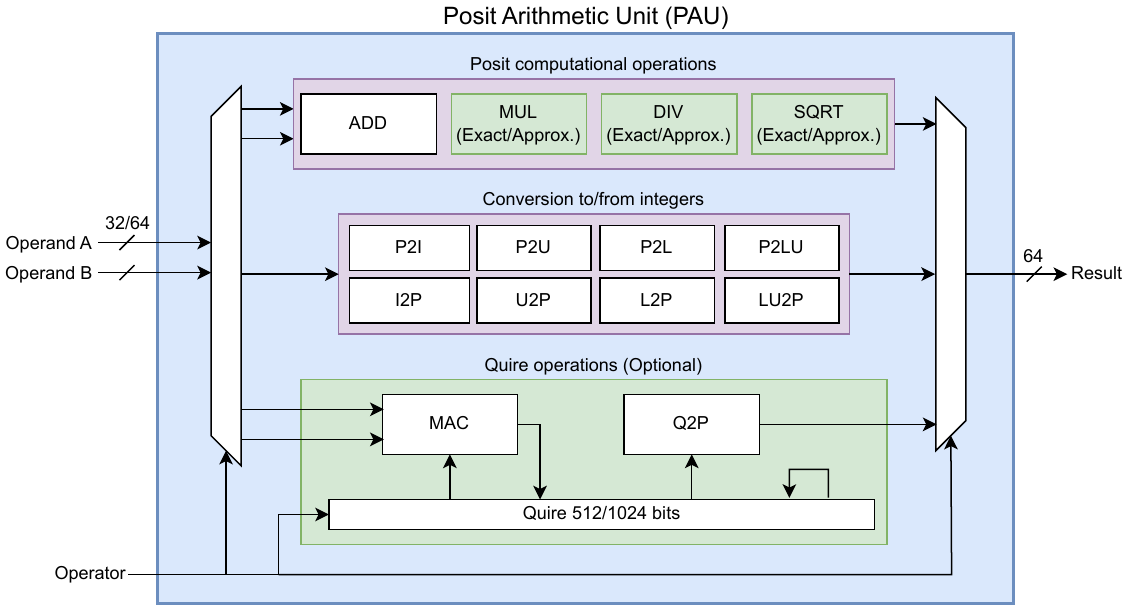}
    \caption{Internal structure of the \acrfull{PAU}. The modifiable blocks with SystemVerilog parameters are shown in green.}
    \label{fig:pau}
\end{figure*}

In addition to adding 64-bit versions of all the computational and conversion units, we include some additional customizations to the \gls{PAU} (see Figure~\ref{fig:pau}). These are set with parameters in the \gls{RTL} description of the core, so they take effect at synthesis time. Furthermore, as typically the multiplicative arithmetic units are among the most power-hungry modules \cite{liu2012power,barrio2016partial,kim2019efficient}, we permit either the use of the logarithm-approximate units presented in~\cite{murillo2023PLAUs} or exact units for the multiplication, division and square root operations~\revisedChangesTwo{\cite{murillo2023Suitea}}. The exact units for these last two operations use a non-restoring division algorithm \cite{kihwan2012modified}. Also, all of the quire operations are optional and the quire can be enabled or disabled completely. These options allow a fine-grained study of the individual impact of the quire accumulator register and approximate versus exact computational units. The input operands can be either 32 or 64 bits. However, the output will always be 64 bits as the core follows the RV64 architecture (XLEN=64).

The new 64-bit operations were validated with an extensive test suite. These tests cover the full range of instructions that Big-PERCIVAL can execute. To obtain the expected outputs of each series of operations we used the Universal Numbers software library~\cite{omtzigt2020Universal}. 

To compile the new 64-bit instructions we updated the modified LLVM compiler with support for the Xposit RISC\nobreakdash-V custom extension~\cite{mallasen2022PERCIVAL} so that it can generate double-precision loads and stores. The rest of the instructions are maintained as in the original proposal since we only support either posit32 or posit64 at any one time and the decoding can be reused. The updated list of the full Xposit instruction set can be found in Table~\ref{tab:xposit_instructions}.

\begin{table}
\caption{Updated instruction set of the XPosit custom RISC-V extension.}
\label{tab:xposit_instructions}
\centering
\resizebox{\columnwidth}{!}{%
\begin{tabular}{@{}lclclclcccc@{}}
&
\multicolumn{1}{l}{\instbit{31}} &
 &
 &
 &
 &
\multicolumn{1}{r}{\instbit{20}} &
\instbitrange{19}{15} &
\instbitrange{14}{12} &
\instbitrange{11}{7} &
\instbitrange{6}{0} \\
\cline{2-11}
\toprule
PLW       & \multicolumn{6}{c}{imm{[}11:0{]}}                                           & rs1 & 0x1 & rd           & 0x0B \\
PLD       & \multicolumn{6}{c}{imm{[}11:0{]}}                                           & rs1 & 0x5 & rd           & 0x0B \\ 
&
\multicolumn{1}{l}{\instbit{31}} &
 &
&
&
\multicolumn{1}{l}{\instbit{24}} &
\multicolumn{1}{r}{\instbit{20}} &
\instbitrange{19}{15} &
\instbitrange{14}{12} &
\instbitrange{11}{7} &
\instbitrange{6}{0} \\ \midrule
PSW       & \multicolumn{4}{c}{imm{[}11:5{]}}                 & \multicolumn{2}{c}{rs2} & rs1 & 0x3 & imm{[}4:0{]} & 0x0B \\
PSD       & \multicolumn{4}{c}{imm{[}11:5{]}}                 & \multicolumn{2}{c}{rs2} & rs1 & 0x6 & imm{[}4:0{]} & 0x0B \\ 
&
\multicolumn{1}{l}{\instbit{31}} &
\multicolumn{1}{r}{\instbit{27}} &
\multicolumn{1}{r}{\hspace{0.75em}\instbit{26}} &
\multicolumn{1}{l}{\hspace{-1.25em}\instbit{25}} &
\multicolumn{1}{r}{\instbit{24}} &
\multicolumn{1}{l}{\instbit{20}} &
\instbitrange{19}{15} &
\instbitrange{14}{12} &
\instbitrange{11}{7} &
\instbitrange{6}{0} \\ \midrule
PADD.S    & \multicolumn{2}{c}{0x00} & \multicolumn{2}{c}{0x2} & \multicolumn{2}{c}{rs2} & rs1 & 0x0 & rd           & 0x0B \\
PSUB.S    & \multicolumn{2}{c}{0x01} & \multicolumn{2}{c}{0x2} & \multicolumn{2}{c}{rs2} & rs1 & 0x0 & rd           & 0x0B \\
PMUL.S    & \multicolumn{2}{c}{0x02} & \multicolumn{2}{c}{0x2} & \multicolumn{2}{c}{rs2} & rs1 & 0x0 & rd           & 0x0B \\
PDIV.S    & \multicolumn{2}{c}{0x03} & \multicolumn{2}{c}{0x2} & \multicolumn{2}{c}{rs2} & rs1 & 0x0 & rd           & 0x0B \\
PMIN.S    & \multicolumn{2}{c}{0x04} & \multicolumn{2}{c}{0x2} & \multicolumn{2}{c}{rs2} & rs1 & 0x0 & rd           & 0x0B \\
PMAX.S    & \multicolumn{2}{c}{0x05} & \multicolumn{2}{c}{0x2} & \multicolumn{2}{c}{rs2} & rs1 & 0x0 & rd           & 0x0B \\
PSQRT.S   & \multicolumn{2}{c}{0x06} & \multicolumn{2}{c}{0x2} & \multicolumn{2}{c}{0x0}   & rs1 & 0x0 & rd           & 0x0B \\
QMADD.S   & \multicolumn{2}{c}{0x07} & \multicolumn{2}{c}{0x2} & \multicolumn{2}{c}{rs2} & rs1 & 0x0 & 0x0            & 0x0B \\
QMSUB.S   & \multicolumn{2}{c}{0x08} & \multicolumn{2}{c}{0x2} & \multicolumn{2}{c}{rs2} & rs1 & 0x0 & 0x0            & 0x0B \\
QCLR.S    & \multicolumn{2}{c}{0x09} & \multicolumn{2}{c}{0x2} & \multicolumn{2}{c}{0x0}   & 0x0   & 0x0 & 0x0            & 0x0B \\
QNEG.S    & \multicolumn{2}{c}{0x0A} & \multicolumn{2}{c}{0x2} & \multicolumn{2}{c}{0x0}   & 0x0   & 0x0 & 0x0            & 0x0B \\
QROUND.S  & \multicolumn{2}{c}{0x0B} & \multicolumn{2}{c}{0x2} & \multicolumn{2}{c}{0x0}   & 0x0   & 0x0 & rd           & 0x0B \\
PCVT.W.S  & \multicolumn{2}{c}{0x0C} & \multicolumn{2}{c}{0x2} & \multicolumn{2}{c}{0x0}   & rs1 & 0x0 & rd           & 0x0B \\
PCVT.WU.S & \multicolumn{2}{c}{0x0D} & \multicolumn{2}{c}{0x2} & \multicolumn{2}{c}{0x0}   & rs1 & 0x0 & rd           & 0x0B \\
PCVT.L.S  & \multicolumn{2}{c}{0x0E} & \multicolumn{2}{c}{0x2} & \multicolumn{2}{c}{0x0}   & rs1 & 0x0 & rd           & 0x0B \\
PCVT.LU.S & \multicolumn{2}{c}{0x0F} & \multicolumn{2}{c}{0x2} & \multicolumn{2}{c}{0x0}   & rs1 & 0x0 & rd           & 0x0B \\
PCVT.S.W  & \multicolumn{2}{c}{0x10} & \multicolumn{2}{c}{0x2} & \multicolumn{2}{c}{0x0}   & rs1 & 0x0 & rd           & 0x0B \\
PCVT.S.WU & \multicolumn{2}{c}{0x11} & \multicolumn{2}{c}{0x2} & \multicolumn{2}{c}{0x0}   & rs1 & 0x0 & rd           & 0x0B \\
PCVT.S.L  & \multicolumn{2}{c}{0x12} & \multicolumn{2}{c}{0x2} & \multicolumn{2}{c}{0x0}   & rs1 & 0x0 & rd           & 0x0B \\
PCVT.S.LU & \multicolumn{2}{c}{0x13} & \multicolumn{2}{c}{0x2} & \multicolumn{2}{c}{0x0}   & rs1 & 0x0 & rd           & 0x0B \\
PSGNJ.S   & \multicolumn{2}{c}{0x14} & \multicolumn{2}{c}{0x2} & \multicolumn{2}{c}{rs2} & rs1 & 0x0 & rd           & 0x0B \\
PSGNJN.S  & \multicolumn{2}{c}{0x15} & \multicolumn{2}{c}{0x2} & \multicolumn{2}{c}{rs2} & rs1 & 0x0 & rd           & 0x0B \\
PSGNJX.S  & \multicolumn{2}{c}{0x16} & \multicolumn{2}{c}{0x2} & \multicolumn{2}{c}{rs2} & rs1 & 0x0 & rd           & 0x0B \\
PMV.X.W   & \multicolumn{2}{c}{0x17} & \multicolumn{2}{c}{0x2} & \multicolumn{2}{c}{0x0}   & rs1 & 0x0 & rd           & 0x0B \\
PMV.W.X   & \multicolumn{2}{c}{0x18} & \multicolumn{2}{c}{0x2} & \multicolumn{2}{c}{0x0}   & rs1 & 0x0 & rd           & 0x0B \\
PEQ.S     & \multicolumn{2}{c}{0x19} & \multicolumn{2}{c}{0x2} & \multicolumn{2}{c}{rs2} & rs1 & 0x0 & rd           & 0x0B \\
PLT.S     & \multicolumn{2}{c}{0x1A} & \multicolumn{2}{c}{0x2} & \multicolumn{2}{c}{rs2} & rs1 & 0x0 & rd           & 0x0B \\
PLE.S     & \multicolumn{2}{c}{0x1B} & \multicolumn{2}{c}{0x2} & \multicolumn{2}{c}{rs2} & rs1 & 0x0 & rd           & 0x0B \\ \bottomrule
\end{tabular}%
}
\end{table}

\section{\removedChangesTwo{FPGA }Synthesis Results} \label{sec:synthesis_results}

In this section, we present \gls{FPGA} \revisedChangesTwo{ and \gls{ASIC}} synthesis results of different configurations of Big-PERCIVAL as well as detailed area costs of the individual posit units. This highlights the hardware cost of posit numbers and the quire, which is the main drawback we have seen. \removedChangesTwo{In every case, the target frequency of 50 MHz was met. This parameter is defined by the base CVA6 core.}

\subsection{\revisedChangesTwo{FPGA}}

For the \gls{FPGA} synthesis results we used Vivado v.2021.2 targeting a Genesys II (Xilinx Kintex-7 XC7K325T-2FFG900C) board. \revisedChangesTwo{In every case, the target frequency of 50 MHz was met. This parameter is defined by the base CVA6 core.} With this setup, we synthesized different configurations of the \gls{PAU} with or without quire and with or without approximate division and square-root units. These results are shown in Table~\ref{tab:fpga_synthesis_pau}.

\begin{table}[t!]
\centering
\caption{Area results of the \gls{PAU} for different configurations of Big-PERCIVAL.}
\label{tab:fpga_synthesis_pau}
\begin{tabular}{@{}lll
                    S[table-format=5.0]
                    S[table-format=4.0]
                    S[table-format=2.0]
                    S[table-format=3.0]
                @{}}
\toprule
\multicolumn{2}{l}{PAU}                             & DivSqrt & \text{LUTs}  & \text{FFs}  & \text{DSPs} & \text{SRLs} \\ \midrule
\multirow{4}{*}{32-bit} & \multirow{2}{*}{No quire} & Approx. & 5666  & 689  & 2   & 0   \\
                        &                           & Exact   & 5923  & 1453 & 2   & 36  \\ \cmidrule(l){2-7} 
                        & \multirow{2}{*}{Quire}    & Approx. & 11605 & 2923 & 4   & 0   \\
                        &                           & Exact   & 12908 & 3640 & 4   & 35  \\ \midrule
\multirow{3}{*}{64-bit} & \multirow{2}{*}{No quire} & Approx. & 8561  & 1075 & 12  & 0   \\
                        &                           & Exact   & 15959 & 4233 & 12  & 71  \\ \cmidrule(l){2-7} 
                        & Quire                     & Exact   & 29781 & 7274 & 24  & 584 \\ \bottomrule
\end{tabular}%
\end{table}

When comparing with the \gls{FPU} available in the CVA6 (FPNew~\cite{mach2021FPnew}), we observe that 32-bit posits and especially the 64-bit variant require significantly more resources than the same length floats. The 32-bit \gls{FPU} occupies 4045 \glspl{LUT}, 1008 \glspl{FF} and 2 \gls{DSP} blocks. The 64-bit only variant of the \gls{FPU} (no 32-bit support) requires 6243 \glspl{LUT}, 1893 \glspl{FF} and 9 \gls{DSP} blocks.

The exact \glspl{PAU} without quire need almost 50\% more \glspl{LUT} and \glspl{FF} in the 32-bit case and more than $2.5\times$ as many resources in the 64-bit case. This shows that the growth in terms of hardware usage is significantly higher in the case of posit numbers. Adding quire capabilities results in an even higher hardware cost, but as will be shown in Section~\ref{sec:polybench} this allows for obtaining more accurate results with \removedChanges{ without any performance degradation over }\revisedChanges{the same memory bandwidth} as double-precision floats. 

If the target application can tolerate some errors in the division and square root outputs, using the logarithm-approximate units \cite{murillo2023PLAUs} allows for a significant reduction in \gls{LUT} and \gls{FF} usage. This is especially true in the 64-bit case, where the total \gls{PAU} \glspl{LUT} can be cut in half and the \glspl{FF} reduced to only 25\%.

Detailed area results for the individual units in the 64-bit \gls{PAU} are shown in Table~\ref{tab:fpga_synthesis_units}. \revisedChanges{As a reference, the whole Big-PERCIVAL core requires} 77888 \glspl{LUT}, 33437 \glspl{FF}, and 51 \gls{DSP} blocks. Unsurprisingly, the largest unit corresponds to the quire \gls{MAC} operation which, together with the quire to posit and additional quire logic in the \gls{PAU}, corresponds to 50\% of the hardware resources of the \gls{PAU}. This goes in line with the results for the 32-bit \gls{PAU} with quire. The exact division and square root units are the next largest components of the \gls{PAU}. The cost of these exact units is substantially reduced in the case of using the corresponding approximate units. As the works in \cite{kim2019efficient,ansari2021improved,murillo2021PLAM,murillo2023PLAUs} highlight, error-resilient applications may benefit from those. Such is the case of \glspl{DNN}, filters, and other Machine Learning kernels.

\begin{table}[t!]
\centering
\caption{Area results of the 64-bit components of the \gls{PAU}.}
\label{tab:fpga_synthesis_units}
\begin{tabular}{@{}l
                   S[table-format=4.0]
                   S[table-format=4.0]
                   S[table-format=2.0]
                   S[table-format=3.0]
                @{}}
\toprule
Name               & \text{LUTs}   & \text{FFs}   & \text{DSPs} & \text{SRLs} \\ \midrule
Quire MAC          & 9834   & 1909  & 12  & 515 \\
Posit Div          & 5710   & 1903  & 0   & 61  \\
Posit Sqrt         & 3447   & 1442  & 0   & 10  \\
Quire to Posit     & 3290   & 236   & 0   & 0   \\
Posit Add          & 1544   & 77    & 0   & 0   \\
Posit Mult         & 1503   & 114   & 12  & 0   \\
Posit Div Approx.  & 870    & 106   & 0   & 0   \\
Posit Sqrt Approx. & 618    & 68    & 0   & 0   \\
Posit to ULong     & 595    & 0     & 0   & 0   \\ 
Posit to Long      & 580    & 0     & 0   & 0   \\
ULong to Posit     & 500    & 0     & 0   & 0   \\
Long to Posit      & 429    & 0     & 0   & 0   \\
UInt to Posit      & 356    & 0     & 0   & 0   \\
Posit to UInt      & 342    & 0     & 0   & 0   \\
Posit to Int       & 311    & 0     & 0   & 0   \\
Int to Posit       & 202    & 0     & 0   & 0   \\ \bottomrule
\end{tabular}
\end{table}

\subsection{\revisedChangesTwo{ASIC}}

\revisedChangesTwo{Regarding \gls{ASIC} synthesis we targeted TSMC's 28nm standard-cell library to obtain more insight into the area and power cost of the 64-bit PAU and FPU. The synthesis was performed using Synopsys DC with a 5ns timing constraint and a toggle rate of 0.1.}

\revisedChangesTwo{The 64-bit FPU (without 32-bit support enabled to match the PAU) requires an area of 21853}~$\mu\text{m}^2$\revisedChangesTwo{ and consumes 0.738~mW of power. On the other hand, the exact 64-bit PAU with quire occupies 114695}~$\mu\text{m}^2$\revisedChangesTwo{ and consumes 3.516~mW of power. In the case of the 64-bit PAU without quire, this results in significantly less resource use. In particular it spans 71090}~$\mu\text{m}^2$\revisedChangesTwo{ and consumes 1.958~mW. These ASIC area and power results go in line with the FPGA results we obtained previously, highlighting the main drawback of current hardware implementations of posit arithmetic.
}

\section{PolyBench} \label{sec:polybench}

In this work, we used the PolyBench suite~\cite{pouchet2016PolyBench} to benchmark Big\nobreakdash-PERCIVAL. This benchmark suite contains a series of numerical computations with static control flow from various domains such as linear algebra computations or physics simulation. From these, we selected some representative algorithms to study how \revisedChanges{posit32 and} posit64 compare to IEEE 754 \revisedChanges{floats and} doubles in scientific computing calculations.

PolyBench implements each benchmark in a single file, with some header parameters and a series of compile-time directives. In order to compile them we have employed the modified version of LLVM with Xposit. Furthermore, the compile-time directives allow both the study of the accuracy of the results and the accurate measuring of the performance using cache flushing and multiple executions.

We have chosen to port to posit arithmetic the following representative benchmarks:
\begin{itemize}
    \item Covariance: Computes the covariance of $N$ data points, each with $M$ attributes.
    \item GEMM: Generalized matrix multiply from BLAS~\cite{dongarra1990Set} $C_{out}=\alpha AB + \beta C$.
    \item 3mm: Linear algebra kernel that consists of three matrix multiplications $G=(AB)(CD)$.
    \item Cholesky: Cholesky decomposition of a positive-definite matrix $A$ into a lower triangular matrix $L$ such that $A=LL^T$. 
    \item Durbin: This is an algorithm for solving Yule-Walker equations, which are a special case of Toeplitz systems.
    \item ludcmp: LU decomposition followed by forward and backward substitutions to solve a system of linear equations.
    \item fdtd-2d: Simplified Finite-Difference Time-Domain method for 2D data. This models electric and magnetic fields based on Maxwell's equations.
    \item seidel-2d: Gauss-Seidel style computation over 2D data with a 9-point stencil pattern.
\end{itemize}

Porting these benchmarks entailed translating the arithmetic kernels from using IEEE 754 \revisedChanges{floats or} doubles to inline assembly with \revisedChanges{posit32 or} posit64 instructions. The code structure was kept unchanged, including the initialization phase which populated the input data to the algorithms. All tests were compiled with the -O3 flag of gcc and clang to optimize the execution of the kernels and minimize the difference between the original code and the straightforward translation to posit assembly.

For each configuration, we provide \revisedChanges{six} executions: \revisedChanges{single- and} double-precision IEEE 754 arithmetic with fused \gls{MAC} operations as optimized by the compiler, \revisedChanges{posit32 and} posit64 arithmetic with quire fused \gls{MAC} operations, and \revisedChanges{posit32 and} posit64 arithmetic without quire, that is, replacing these \gls{MAC} operations with individual multiplication and addition operations. The fdtd-2d and seidel-2d benchmarks do not benefit from fused \gls{MAC} operations, hence they have a single \revisedChanges{posit32 and} posit64 execution.

Each of these benchmarks was executed with four problem sizes. These range from the MINI datasets, which require less than 16KB of memory each, to the LARGE datasets, which occupy around 25MB of memory. The cache hierarchy of PERCIVAL comprises a 32KB 8-way set-associative L1 data cache with 16-byte lines. Therefore, these problem sizes stretch the whole memory range.

The performance results as shown by the timing script of PolyBench are shown in Table~\ref{tab:polyb_time}. As can be seen, \revisedChanges{float is faster than posit32 in practically all scenarios}. \revisedChanges{However,} when comparing doubles and posit64, there is no clear winner except for the GEMM benchmark, which will be studied in detail in Section~\ref{sec:gemm}. In the rest of the algorithms, some perform better using doubles and others using posit64. These variations are due in part to the translation of the execution kernels from doubles in the native C datatype to the posit64 inline assembly of the Xposit RISC-V custom extension. Furthermore, the execution of 64-bit posits without quire is slower than the other options, as an individual multiplication plus addition has more latency than a single fused \gls{MAC} operation both using doubles or posit64 with quire. Again, this does not hold in the GEMM benchmark, which will be detailed in Section~\ref{sec:gemm}. Therefore we can conclude that\revisedChanges{, when integrating} 64-bit posits with quire \revisedChanges{into the CVA6, they} do not suffer any noticeable performance degradation in comparison with double-precision floats. \revisedChanges{This system inherits the limitations given by the underlying core, and since the critical path is not in the PAU or the FPU, but in the load/store unit, further optimizations are out of the scope of this work.}

\begin{table}[!t]
\centering
\caption{PolyBench timing comparisons of \revisedChanges{floats and} doubles with fused MAC, \revisedChanges{posit32 and} posit64 with quire, and \revisedChanges{posit32 and} posit64 without quire (\revisedChanges{posit32-- and} posit64--) measured in seconds of \gls{FPGA} runtime.}
\label{tab:polyb_time}
\resizebox{\columnwidth}{!}{%
\begin{tabular}{@{}llcccc@{}}
\toprule
                            &           & MINI     & SMALL  & MEDIUM & LARGE  \\ \midrule
\multirow{3}{*}{GEMM}       & float     & 0.004991 & 0.1004 & 4.579  &  591.8 \\
                            & posit32   & 0.005175 & 0.1151 & 5.225  & 1251.3 \\
                            & posit32-- & 0.007326 & 0.1638 & 6.382  &  846.5 \\
                            & double    & 0.006025 & 0.1846 & 6.988  & 1004.9 \\
                            & posit64   & 0.005845 & 0.1747 & 7.500  & 1973.3 \\
                            & posit64-- & 0.007628 & 0.2175 & 7.981  & 1109.5 \\ \midrule
\multirow{3}{*}{3mm}        & float     & 0.005591 & 0.1177 & 9.533  & 2208.0 \\
                            & posit32   & 0.003778 & 0.1250 & 9.678  & 2265.2 \\
                            & posit32-- & 0.009895 & 0.2230 & 14.01  & 2757.0 \\
                            & double    & 0.005321 & 0.2307 & 16.09  & 3931.2 \\
                            & posit64   & 0.004626 & 0.1903 & 14.14  & 3674.6 \\
                            & posit64-- & 0.010466 & 0.3044 & 18.46  & 4208.8 \\ \midrule
\multirow{3}{*}{Cholesky}   & float     & 0.002504 & 0.0617 & 3.670  & 484.0 \\
                            & posit32   & 0.003005 & 0.0707 & 3.811  & 498.5 \\
                            & posit32-- & 0.004960 & 0.1179 & 5.699  & 734.8 \\
                            & double    & 0.003952 & 0.1190 & 6.323  & 870.0 \\
                            & posit64   & 0.003453 & 0.0987 & 5.529  & 767.3 \\
                            & posit64-- & 0.005044 & 0.1503 & 7.680  & 1038.1 \\ \midrule
\multirow{3}{*}{Durbin}     & float     & 0.000526 & 0.00388 & 0.0394  & 1.011 \\
                            & posit32   & 0.001091 & 0.00660 & 0.0676  & 1.681 \\
                            & posit32-- & 0.000814 & 0.00724 & 0.0708  & 1.766 \\
                            & double    & 0.000772 & 0.00703 & 0.0698  & 2.127 \\
                            & posit64   & 0.000898 & 0.00732 & 0.0814  & 2.445 \\
                            & posit64-- & 0.000928 & 0.00816 & 0.0851  & 2.539 \\ \midrule
\multirow{3}{*}{\begin{tabular}[c]{@{}l@{}}LU dcmp\\ solver\end{tabular}} & float     & 0.004377 & 0.1079 & 9.162  & 2896.6 \\
                            & posit32   & 0.006173 & 0.1459 & 10.37  & 2986.3 \\
                            & posit32-- & 0.007162 & 0.1704 & 11.58  & 3147.2 \\
                            & double    & 0.005918 & 0.2447 & 19.48  & 3309.6 \\
                            & posit64   & 0.006388 & 0.2536 & 19.96  & 3311.2 \\
                            & posit64-- & 0.007347 & 0.2859 & 20.47  & 3468.3 \\ \midrule
\multirow{2}{*}{Covariance} & float     & 0.004203 & 0.0939 & 4.381  & 1779.2 \\
                            & posit32   & 0.004447 & 0.0903 & 4.149  & 1727.9 \\
                            & posit32-- & 0.005377 & 0.1062 & 4.552  & 1793.1 \\
                            & double    & 0.004478 & 0.1811 & 7.843  & 1949.5 \\
                            & posit64   & 0.004817 & 0.1670 & 7.311  & 1883.0 \\
                            & posit64-- & 0.005678 & 0.1875 & 7.818  & 1948.8 \\ \midrule
\multirow{2}{*}{FDTD 2D}    & float     & 0.014354 & 0.3824 & 11.14  & 1457.3 \\
                            & posit32   & 0.020327 & 0.5168 & 14.49  & 1893.7 \\
                            & double    & 0.016546 & 0.6750 & 17.37  & 2469.6 \\
                            & posit64   & 0.020668 & 0.7709 & 19.83  & 2711.4 \\ \midrule
\multirow{2}{*}{\begin{tabular}[c]{@{}l@{}}Gauss-Seidel\\ 2D\end{tabular}} & float     & 0.030509 & 0.5979 & 17.35  & 2346.2\\
                            & posit32   & 0.036817 & 0.7519 & 22.74  & 3164.3 \\
                            & double    & 0.041236 & 0.8985 & 26.98  & 4071.4 \\
                            & posit64   & 0.037041 & 0.8412 & 26.55  & 4097.4 \\ \bottomrule
\end{tabular}
}
\end{table}

\begin{figure*}[!t]
    \centering
    \includegraphics[width=\textwidth]{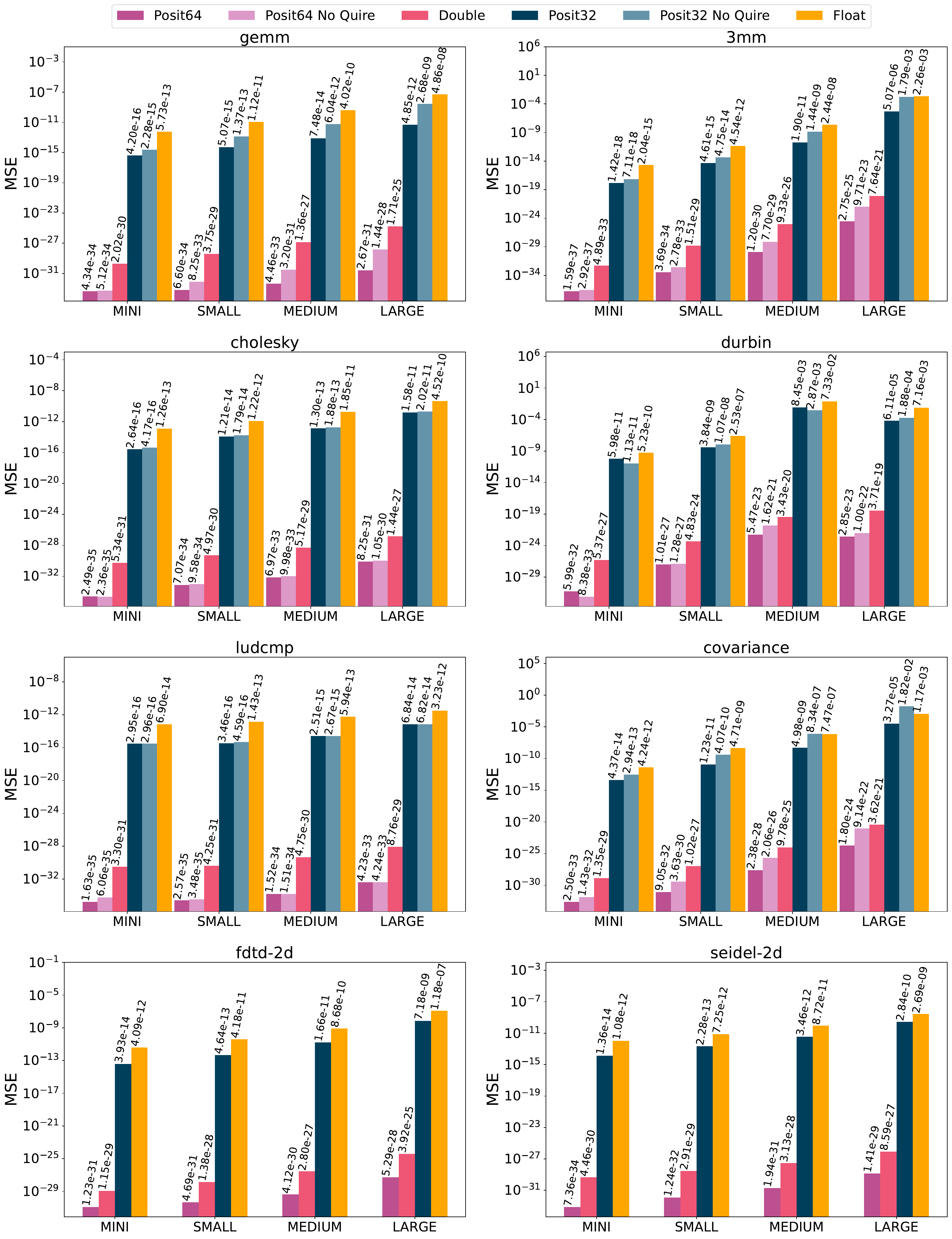}
    \caption{PolyBench benchmarks mean square error of \revisedChanges{the different arithmetics studied} with respect to \revisedChanges{the results obtained with GNU MPFR.}}
    \label{fig:polybench_mse}
\end{figure*}

\begin{figure*}[!t]
    \centering
    \includegraphics[width=\textwidth]{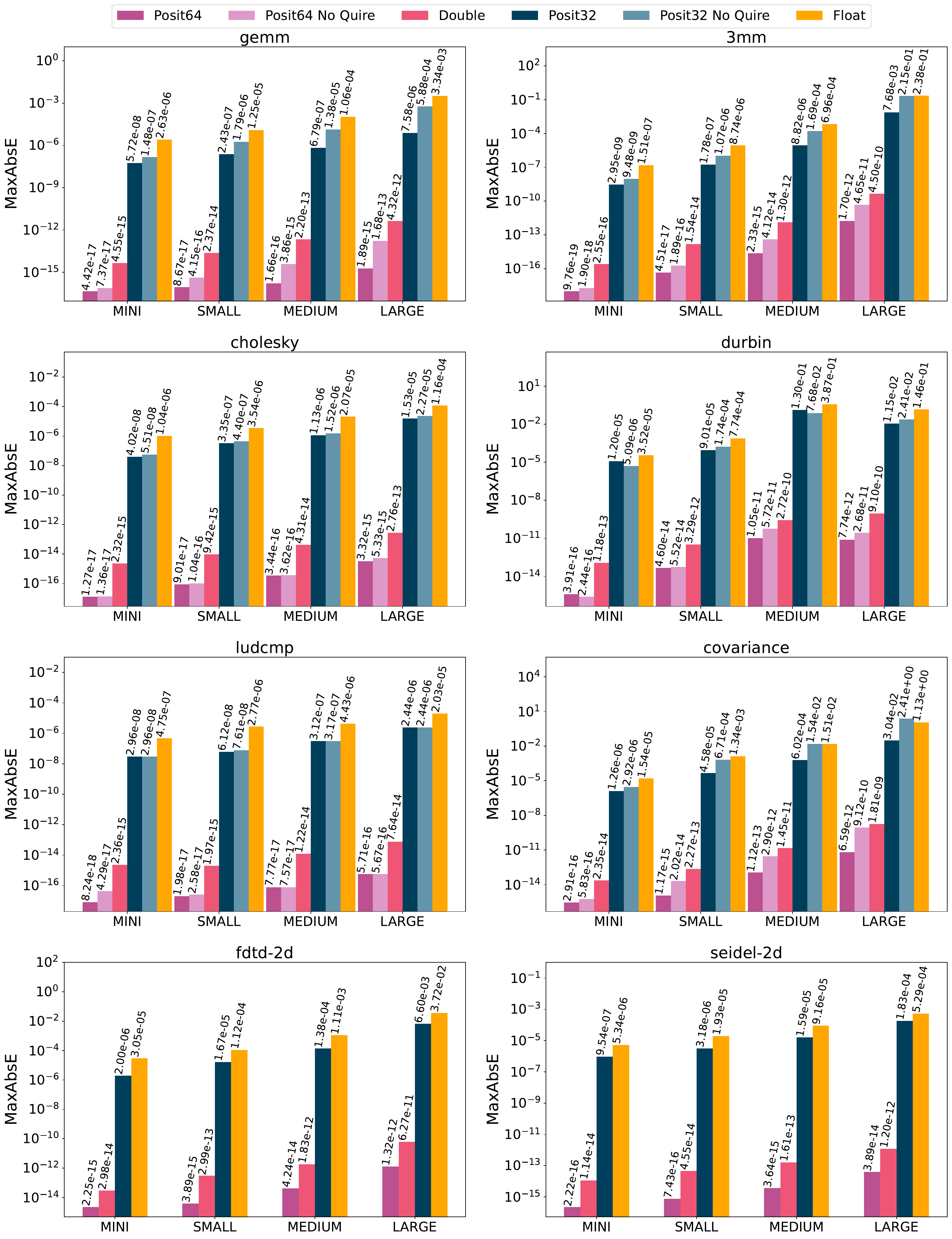}
    \caption{PolyBench benchmarks maximum absolute error of the different arithmetics studied with respect to \revisedChanges{the results obtained with GNU MPFR.}}
    \label{fig:polybench_maxabse}
\end{figure*}

Regarding accuracy, we have obtained two metrics: \acrfull{MSE} and \acrfull{MaxAbsE}. To obtain both of these metrics, we compared the results of \revisedChanges{all the arithmetics under study} to the same algorithm computed using \removedChanges{ double extended 80 bit format present in x86 processors }\revisedChanges{the GNU MPFR multiple-precision library with 128 fraction bits, which we use as our golden solution.} We chose the \gls{MSE} as a general accuracy metric and the \gls{MaxAbsE} to also take into account the maximum error, which can be a critical value in certain applications \cite{xitong2013soap,villalba2018unbiased}.

The \gls{MSE} results are shown in Figure~\ref{fig:polybench_mse}. Note the logarithmic scale on the Y-axis. The trend in every benchmark is a significantly lower error when using posit64 \revisedChanges{or posit32} numbers \revisedChanges{in comparison to floats or doubles, respectively.} This is up to 4 orders of magnitude lower \gls{MSE} depending on the benchmark. \revisedChanges{In these plots we can also observe the difference in magnitude of the accuracy errors when using 32- or 64-bit numbers.}

The accuracy improvements are maintained across the whole range of problem sizes. When using posits without the quire accumulator we also observe significant accuracy improvements. This happens even though in the posit case we have an extra rounding between the multiplication and addition operations that is not present when using fused \gls{MAC} with \removedChanges{ doubles }\revisedChanges{IEEE representations.}

The \gls{MaxAbsE} metric follows the same pattern as the \gls{MSE}. These results are shown in Figure~\ref{fig:polybench_maxabse}, where the Y-axis also follows a logarithmic scale. \revisedChanges{Posits} obtain up to 3 orders of magnitude lower error than \revisedChanges{their same bit-width floats} across all benchmarks and dataset sizes. There are also large accuracy improvements when executing with posit without quire over doubles.

All in all, from this set of benchmarks we can conclude that 64-bit posits\removedChanges{ can execute as fast as double-precision floats. Moreover, posit64 } present significant accuracy improvements over IEEE 754 doubles, while maintaining the same memory bandwidth. \revisedChanges{This is also the case with 32-bit numbers.} This better accuracy is maintained both in a general sense, as shown by the \gls{MSE}, as well as in each particular value since the \gls{MaxAbsE} is also lower in every case. The use of the quire is beneficial both in terms of performance and accuracy, so it can compensate for its significant hardware cost shown in Section~\ref{sec:synthesis_results}.

\section{Conjugate Gradient} \label{sec:cg}

\revisedChanges{In addition to the Polybench algorithms shown in} Section~\ref{sec:polybench}, \revisedChanges{we have studied the use of posit64 in iterative linear equation solvers. In particular, when using the conjugate gradient (CG) and the biconjugate gradient (BiCG) methods. These serve as larger real-world applications in which posit64 could be used.}

\revisedChanges{The conjugate gradient algorithm serves to numerically solve systems of linear equations} $\mathbf{Ax} = \mathbf{b}$ \revisedChanges{for vector} $\mathbf{x}$, \revisedChanges{where the real matrix} $\mathbf{A}$ \revisedChanges{is symmetric and positive-definite. We have executed this algorithm on Big-PERCIVAL with a tolerance margin of} $10^{-12}$ \revisedChanges{on four matrices extracted from the Matrix Market repository}\footnote{\revisedChanges{https://math.nist.gov/MatrixMarket/}}. \revisedChanges{Concretely, on a subset of the BCS Structural Engineering Matrices from the Harwell-Boeing Collection. This provided a real use-case in which we can analyze the use of posit64 and compare it with IEEE doubles.}

\begin{figure*}
    \centering
    \includegraphics[width=0.875\textwidth]{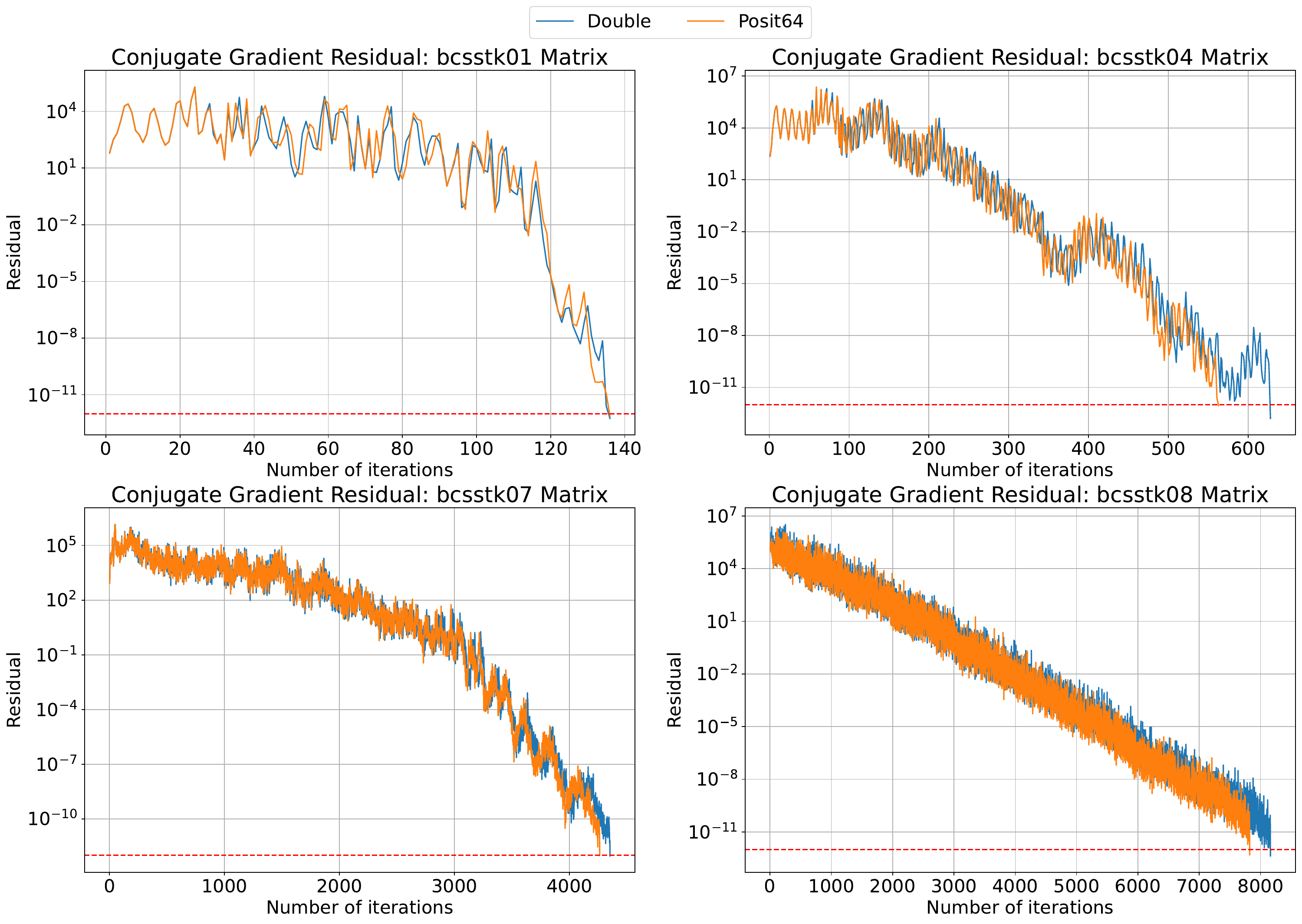}
    \caption{\revisedChanges{Conjugate gradient iterative residual results on four increasingly larger Matrix Market problems.}}
    \label{fig:cg_residual}
\end{figure*}

Figure~\ref{fig:cg_residual} \revisedChanges{shows the results of these executions. The bcsstk01, bcsstk04, bcsstk07, and bcsstk08 matrices have a size of} $48\times 48$, $132\times 132$, $420\times 420$, and $1074\times 1074$, \revisedChanges{respectively. In the smallest case, both posit64 and double converge in the same number of iterations (136), but in the rest of the problems, posit64 converges in fewer iterations. This amounts to a reduction of} 2-10\% \revisedChanges{in the number of iterations of the algorithm, depending on the input matrix.}

\begin{figure*}
    \centering
    \includegraphics[width=0.875\textwidth]{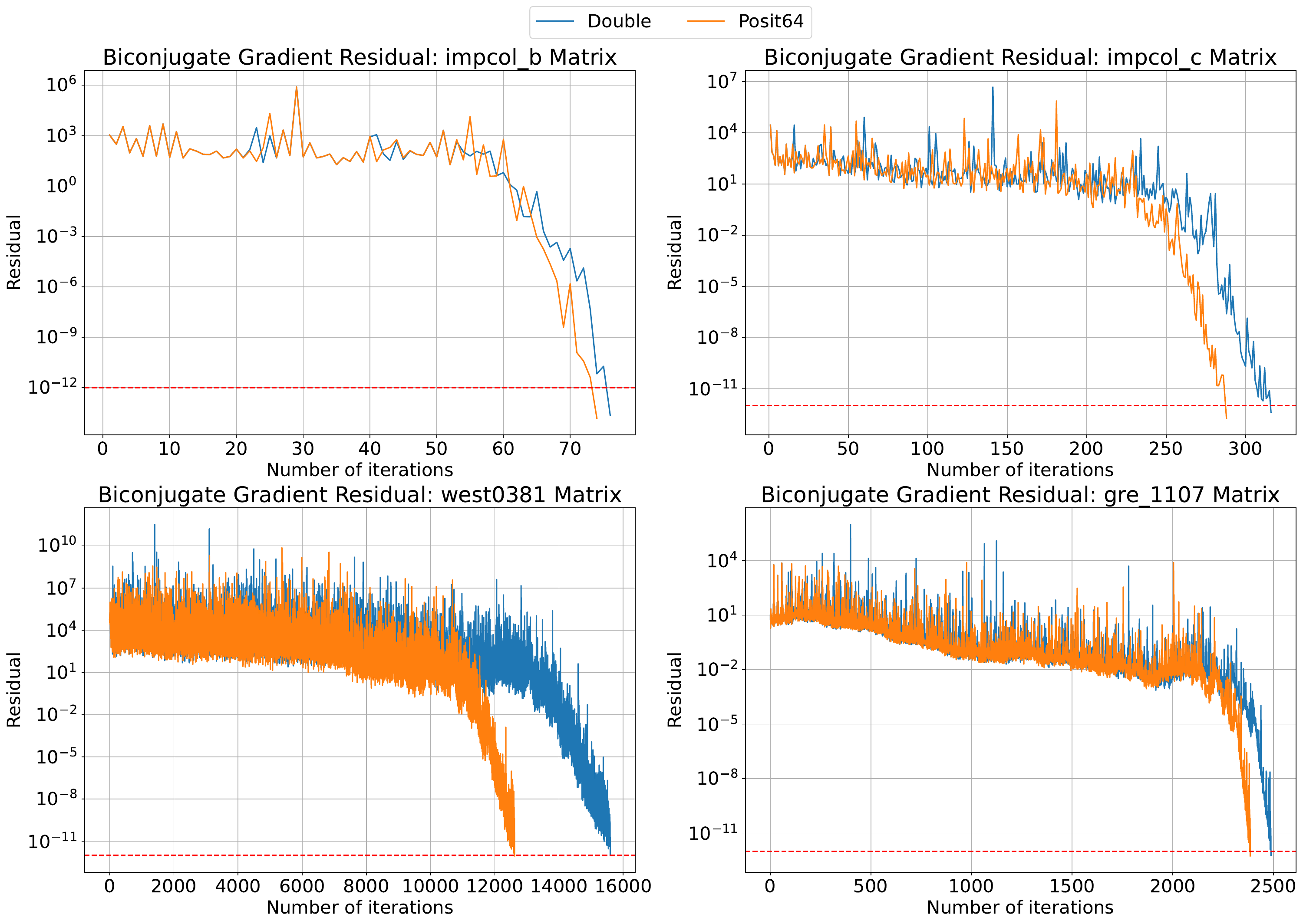}
    \caption{\revisedChanges{Biconjugate gradient iterative residual results on four increasingly larger Matrix Market problems.}}
    \label{fig:bicg_residual}
\end{figure*}

\revisedChanges{The biconjugate gradient method is a generalized version of the CG that serves to solve systems of linear equations in which matrix} $\mathbf{A}$ \revisedChanges{does not have the restriction of being symmetric and positive-definite. On the other hand, its computational cost is around double the one of CG.}

\revisedChanges{Following the same setup, we tested BiCG on a set of four increasingly larger unsymmetric matrices from the Harwell-Boeing Collection. These were the} impcol\_b, impcol\_c, west0381, and gre\_1107, \revisedChanges{which have a size of} $59\times 59$, $137\times 137$, $381\times 381$, and $1107\times 1107$, \revisedChanges{respectively. The tolerance margin was again} $10^{-12}$. \revisedChanges{The results can be seen in} Figure~\ref{fig:bicg_residual}. \revisedChanges{In every case, using posit64 results in fewer iterations before the target tolerance is met. This reduction reaches almost} 20\% \revisedChanges{in the west0381 matrix.}

\section{GEMM} \label{sec:gemm}

The GEMM kernel in PolyBench is optimized from the memory perspective by performing loop interchange (see loops \texttt{k} and \texttt{j} in Figure~\ref{alg:double_gemm}). This version is the one whose results are shown in Section~\ref{sec:polybench}. When executing the GEMM kernel in hardware for sufficiently large matrices it can be observed that there is a considerable time penalty for the posit64 case (see Table~\ref{tab:polyb_time}). This is due to the fact that exploiting the use of the quire accumulator register limits the order in which matrix multiplication is computed (see Figure~\ref{alg:posit_gemm}), which results in a higher number of cache misses because of the long dot-product computations. This is not the case for the execution without the quire, where the performance results are comparable.

\begin{figure}
\begin{algorithmic}
    \REQUIRE Double matrices \texttt{A} (\texttt{ni}$\times$\texttt{nk}), \texttt{B} (\texttt{nk}$\times$\texttt{nj}) and \texttt{C} (\texttt{ni}$\times$\texttt{nj}). Scalar values \texttt{a} and \texttt{b}.
    \ENSURE Double matrix \texttt{C} = \texttt{aAB} + \texttt{bC}.
    \FOR{i = 0 \TO \texttt{ni}-1}
        \FOR{j = 0 \TO \texttt{nj}-1}
            \STATE \texttt{C}[i][j] *= \texttt{b}
        \ENDFOR
        \FOR{k = 0 \TO \texttt{nk}-1}
            \FOR{j = 0 \TO \texttt{nj}-1}
                \STATE \texttt{C}[i][j] += \texttt{a} * \texttt{A}[i][k] * \texttt{B}[k][j]
            \ENDFOR
        \ENDFOR
    \ENDFOR
\end{algorithmic}
\caption{PolyBench Double GEMM pseudocode with loop interchange.}
\label{alg:double_gemm}
\end{figure}

\begin{figure}
\begin{algorithmic}
    \REQUIRE Posit64 matrices \texttt{A} (\texttt{ni}$\times$\texttt{nk}), \texttt{B} (\texttt{nk}$\times$\texttt{nj}) and \texttt{C} (\texttt{ni}$\times$\texttt{nj}). Scalar values \texttt{a} and \texttt{b}.
    \ENSURE Posit64 matrix \texttt{C} = \texttt{aAB} + \texttt{bC}.
    \FOR{i = 0 \TO \texttt{ni}-1}
        \FOR{j = 0 \TO \texttt{nj}-1}
            \STATE \texttt{C}[i][j] *= \texttt{b}
        \ENDFOR
        \FOR{j = 0 \TO \texttt{nj}-1}
            \STATE \texttt{quire} = \texttt{C}[i][j]
            \FOR{k = 0 \TO \texttt{nk}-1}
                \STATE \texttt{quire} += \texttt{a} * \texttt{A}[i][k] * \texttt{B}[k][j]
            \ENDFOR
            \STATE \texttt{C}[i][j] = round(\texttt{quire})
        \ENDFOR
    \ENDFOR
\end{algorithmic}
\caption{Posit GEMM pseudocode using the quire accumulator.}
\label{alg:posit_gemm}
\end{figure}

The GEMM operation is typically optimized to reduce the number of memory accesses. In this section, we describe the impact on both timing performance and accuracy of executing the GEMM kernel using posit64 and quire with block tiling optimizations. The matrices are kept the same as in the PolyBench GEMM benchmark, but we modified the algorithm to use a standard 6-loop tiling approach (see Figure~\ref{alg:posit_gemm_tiled}). The tile size was varied at compile time to better study the impact of this method, allowing the compiler to perform optimizations. We tested all tile sizes between 5 and 25 and also larger tiles of 30 to 40 in steps of 2 to check the observed trends.

\begin{figure}
\begin{algorithmic}
    \REQUIRE Posit64 matrices \texttt{A} (\texttt{ni}$\times$\texttt{nk}), \texttt{B} (\texttt{nk}$\times$\texttt{nj}) and \texttt{C} (\texttt{ni}$\times$\texttt{nj}). Scalar values \texttt{a} and \texttt{b}. Tile size \texttt{nt}.
    \ENSURE Posit64 matrix \texttt{C} = \texttt{aAB} + \texttt{bC}.
    \FOR{ii = 0 \TO \texttt{ni}-1 \textbf{in steps of} \texttt{nt}}
        \FOR{jj = 0 \TO \texttt{nj}-1 \textbf{in steps of} \texttt{nt}}
            \FOR{i = ii \TO \texttt{min}(ii+\texttt{nt}, \texttt{ni})-1}
                \FOR{j = jj \TO \texttt{min}(jj+\texttt{nt}, \texttt{nj})-1}
                    \STATE \texttt{C}[i][j] *= \texttt{b}
                \ENDFOR
            \ENDFOR
        \ENDFOR
        \FOR{kk = 0 \TO \texttt{nk}-1 \textbf{in steps of} \texttt{nt}}
            \FOR{i = ii \TO \texttt{min}(ii+\texttt{nt}, \texttt{ni})-1}
                \FOR{j = jj \TO \texttt{min}(jj+\texttt{nt}, \texttt{nj})-1}
                    \STATE \texttt{quire} = \texttt{C}[i][j]
                    \FOR{k = kk \TO \texttt{min}(kk+\texttt{nt}, \texttt{nk})-1}
                        \STATE \texttt{quire} += \texttt{a} * \texttt{A}[i][k] * \texttt{B}[k][j]
                    \ENDFOR
                    \STATE \texttt{C}[i][j] = round(\texttt{quire})
                \ENDFOR
            \ENDFOR
        \ENDFOR
    \ENDFOR
\end{algorithmic}
\caption{Posit GEMM tiled pseudocode using the quire accumulator.}
\label{alg:posit_gemm_tiled}
\end{figure}

\begin{figure}
    \centering
    \includegraphics[width=\columnwidth]{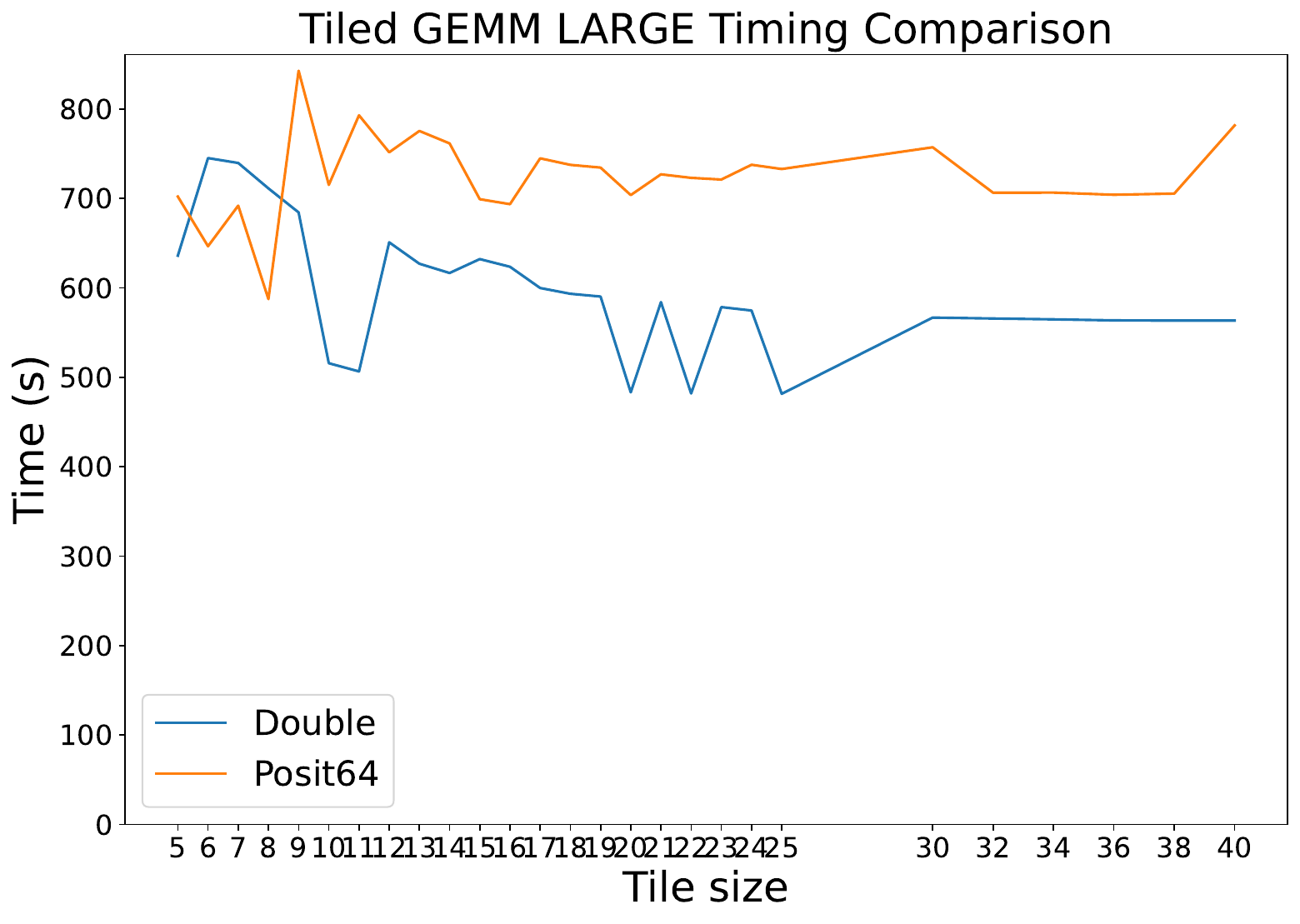}%
    \caption{Tiled GEMM timing results.}
    \label{fig:timing_gemm_tiled}
\end{figure}

\begin{figure*}
    \centering
        \subfloat[]{\includegraphics[width=0.49\textwidth]{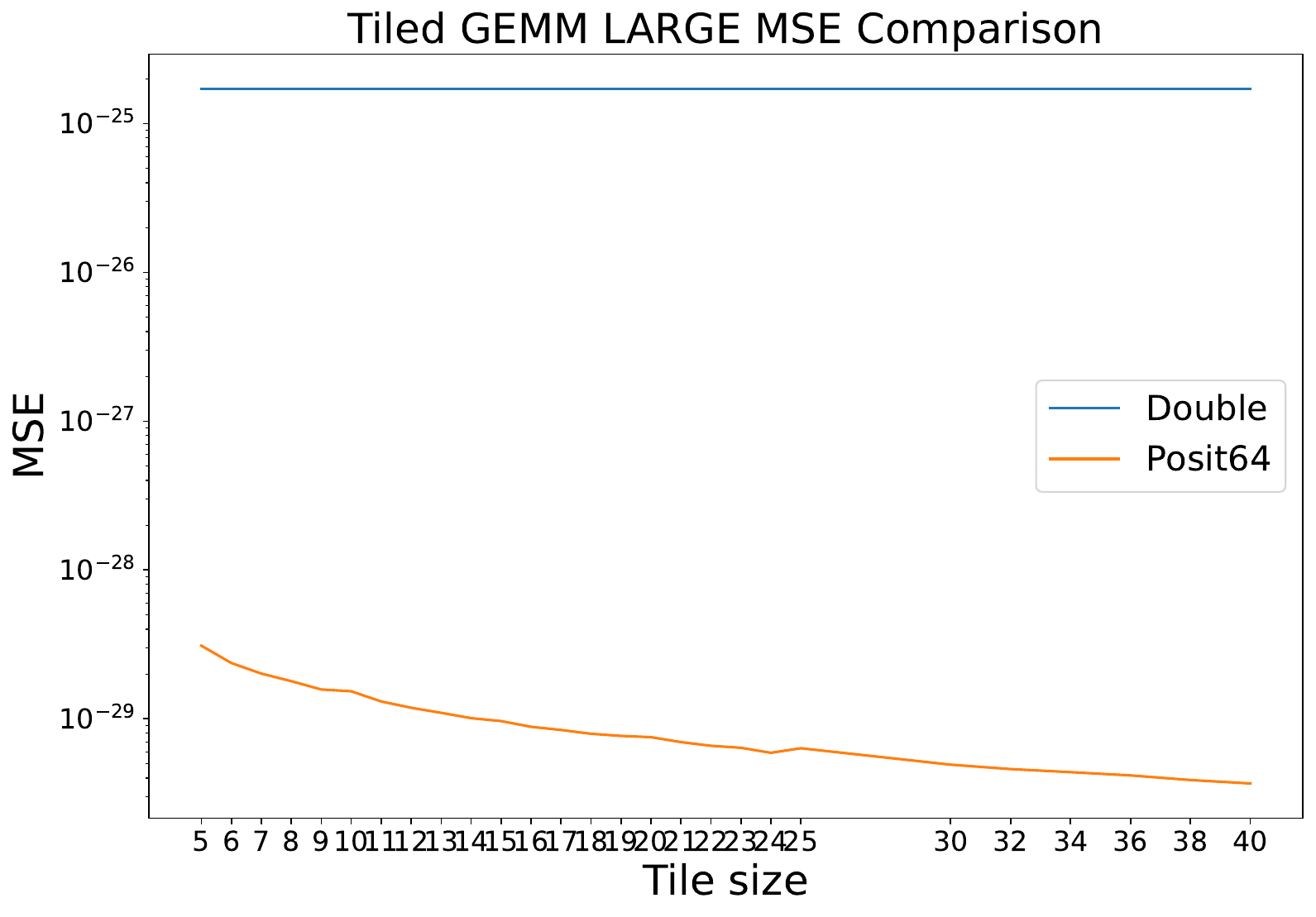}%
        \label{fig:mse_gemm_tiled}}
    \hfill
        \subfloat[]{\includegraphics[width=0.49\textwidth]{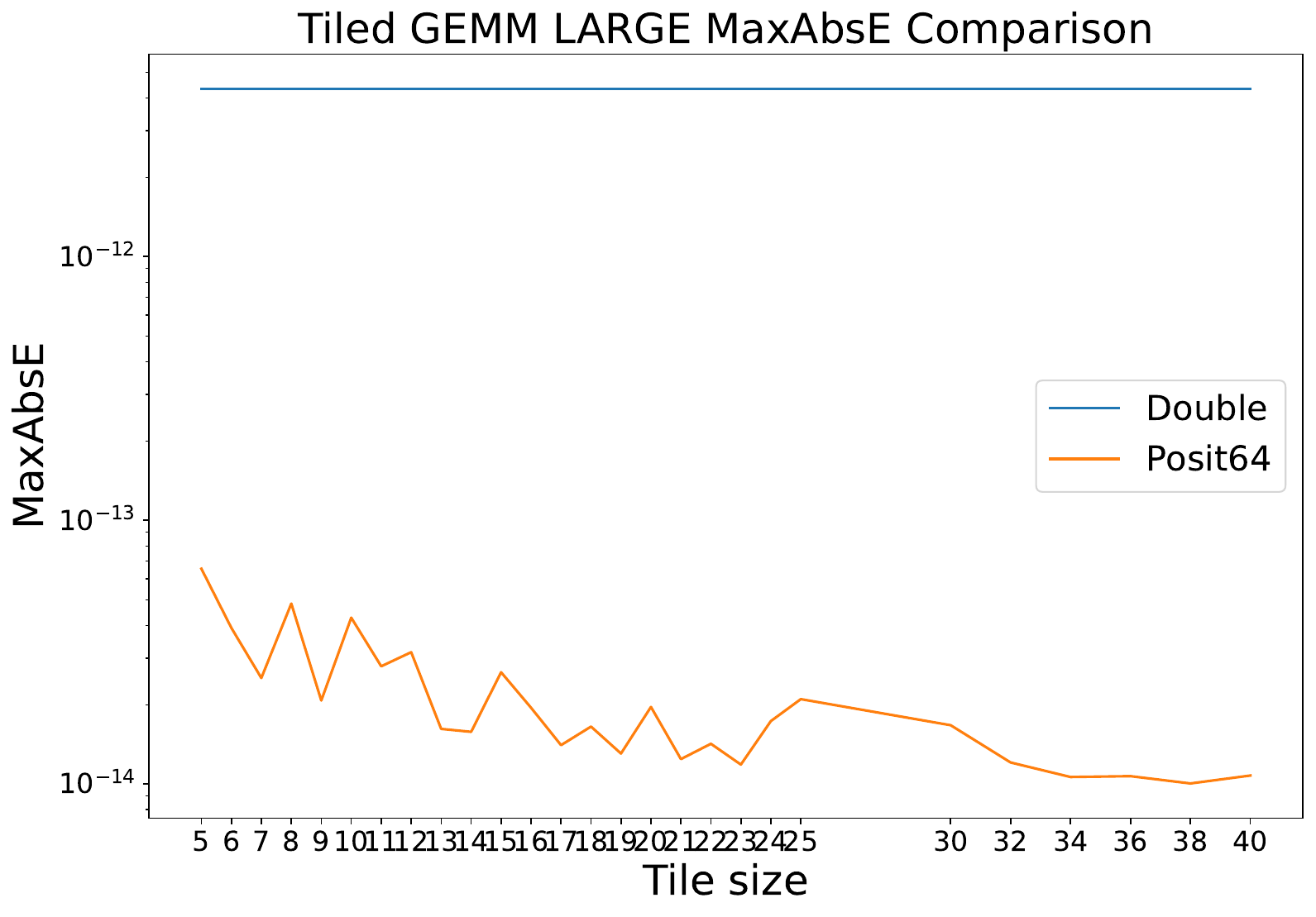}%
        \label{fig:maxabse_gemm_tiled}}
    \caption{Tiled GEMM error comparison of posit64 and double with respect to the results obtained with GNU MPFR.}
    \label{fig:error_gemm_tiled}
\end{figure*}

Figure~\ref{fig:timing_gemm_tiled} shows the timing results of executing the tiled version of GEMM when varying the tile size. These results are for the LARGE dataset, which set the values of \texttt{ni}, \texttt{nj} and \texttt{nk} to $1000$, $1100$ and $1200$ respectively. For very small tile sizes between 5 and 10, there is a relatively large variation in the performance, but this stabilizes for larger tile sizes. In the case of posit64 numbers, there is a big jump in execution time between using a tile size of 8 or a tile size of 9. This is due to compiler optimizations. In the size 8 case, the compiler can loop-unroll the main computational loop and this does not happen with the size 9 case. In the case of doubles, this happens for sizes 9 and 10. When using a tile size of 10 the compiler decides to loop-unroll the main computational loop, and this is not the case for a tile size of 9.

The extra execution time required by the posit64 kernel is due to the extra instructions needed to initialize the quire and round it back to a posit value after each series of accumulations inside a tile. For large dot-product computations, these extra instructions are amortized over the long accumulations, but for smaller batches this overhead is noticeable. All in all, the performance comparison of posit64 and doubles in the GEMM tiled benchmark is closer and should scale better for even larger matrix sizes, as the memory pressure is reduced.

Even though the posit64 execution of this kernel is slower, there are significant benefits regarding the accuracy of the computations. Figure~\ref{fig:error_gemm_tiled} shows the \gls{MSE} and \gls{MaxAbsE} results of the same execution of the GEMM tiled benchmark. As can be seen from the logarithmic scale on the Y-axis, posit64 obtains between 4 and 5 orders of magnitude lower \gls{MSE} and around 2 orders of magnitude lower \gls{MaxAbsE} than doubles. The accuracy improves with larger tile sizes, and is comprised between the posit64 with and without quire values shown in Figures~\ref{fig:polybench_mse} and~\ref{fig:polybench_maxabse}. This is to be expected, as the execution in tiles adds extra rounding steps in the computation of each value of the output matrix. With larger tiles, the number of intermediate roundings will be lower and thus the final value is more accurate. However, note that even for small tile sizes,\removedChanges{ where the posit64 execution time was comparable to the performance of doubles, } the accuracy improvements obtained by posit arithmetic are about 4 orders of magnitude.

\section{Conclusions} \label{sec:conclusions}

In this work, we presented Big-PERCIVAL, an extension of the PERCIVAL posit RISC-V core that adds support for 64-bit posits and provides increased flexibility. We studied the hardware cost, accuracy, and performance of 64-bit posit arithmetic compared to double-precision IEEE 754 floating-point arithmetic using the PolyBench benchmark suite.

Synthesis results of the 64-bit \gls{PAU} in Big-PERCIVAL have shown that it requires $2.5\times$ as many resources as the double-precision FPNew \gls{FPU}. Moreover, we studied the impact of the corresponding 1024-bit quire accumulator register, which increased the total hardware cost to a third of the area of the core. Detailed area results illustrated how the hardware resources are distributed among the different operations. In particular, the most resource-hungry elements are the quire-related units and the posit division and square root units.

The PolyBench numerical benchmarks executed on Big-PERCIVAL running on the Genesys II board provided insight into the native use of 64-bit posits. \removedChanges{ Performance results have shown that posit64 numbers can perform as fast as IEEE 754 double-precision numbers. }\revisedChanges{Furthermore, the conjugate gradient and biconjugate gradient linear solvers demonstrated the use of posit64 in real-world problems.} Additionally, the use of the quire accumulator requires some extra thought into the order in which the operations will be executed in some instances. Regarding accuracy, which is one of the main requirements in scientific computing, we have seen that 64-bit posits obtain up to 4 orders of magnitude lower \gls{MSE} and up to 3 orders of magnitude lower \gls{MaxAbsE} than 64-bit doubles. This provides a high-accuracy solution \revisedChanges{that can reduce the number of steps in iterative solvers} without additional impact on the memory bandwidth.

Overall, our contributions show the potential of posit arithmetic as an alternative to IEEE 754 floating-point arithmetic in scientific computing, and Big-PERCIVAL provides a flexible platform for exploring this alternative. We believe that this work provides a starting point for future research on 64-bit hardware and software implementations of posit arithmetic and contributes to the development of more accurate and efficient scientific computing systems.



\ifCLASSOPTIONcompsoc
  \section*{Acknowledgments}
\else
  \section*{Acknowledgment}
\fi

This work was supported by grants PID2021-123041OB-I00 and PID2021-126576NB-I00 funded by MCIN/AEI/ 10.13039/501100011033 and by “ERDF A way of making Europe”, and by the CM under grant S2018/TCS-4423.

\ifCLASSOPTIONcaptionsoff
  \newpage
\fi



\bibliographystyle{IEEEtran}
\bibliography{references}
%

%

\begin{IEEEbiography}[{\includegraphics[width=1in,height=1.25in,clip,keepaspectratio]{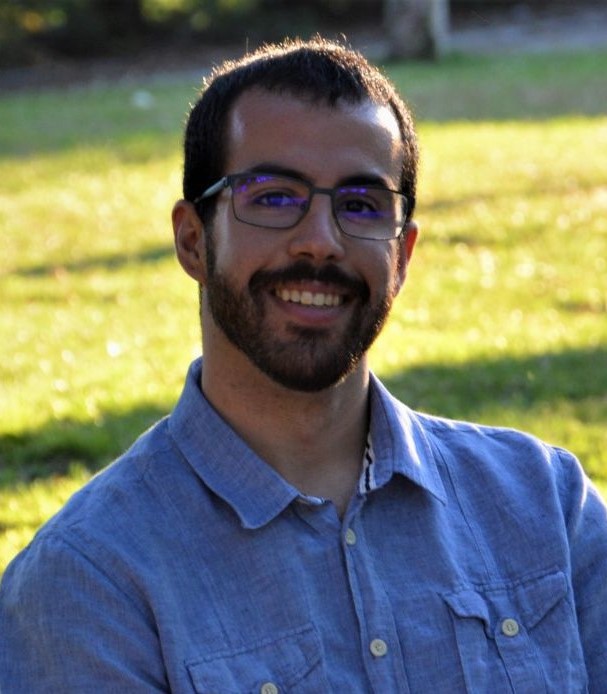}}]{David Mallasén}
David Mallasén Quintana received a BSc Degree in Computer Science and a BSc Degree in Mathematics in 2020 from the Complutense University of Madrid (UCM). In 2022 he obtained an MSc Degree in Embedded Systems at KTH Royal Institute of Technology, specializing in embedded platforms. Currently, he is pursuing a Ph.D. in Computer Engineering at UCM. He has carried out a Ph.D. research stay at the Embedded Systems Laboratory at EPFL (Switzerland). His main research areas include computer arithmetic, computer architecture, embedded systems, and high-performance computing.
\end{IEEEbiography}

\begin{IEEEbiography}[{\includegraphics[width=1in,height=1.25in,clip,keepaspectratio]{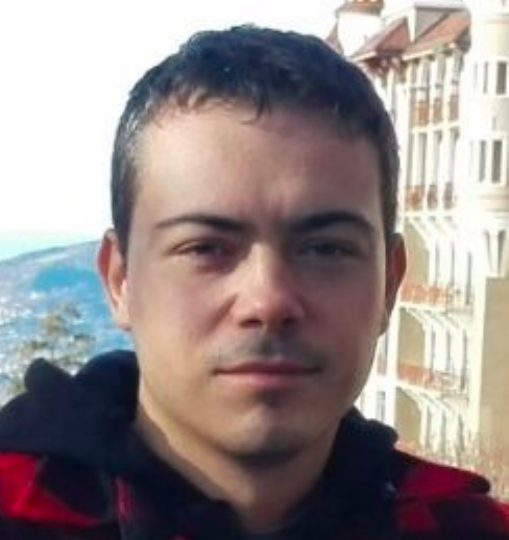}}]{Alberto A. Del Barrio (SM'19)}
Alberto A. Del Barrio received the Ph.D. degree in Computer Science from the Complutense University of Madrid (UCM), Madrid, Spain, in 2011. He has performed stays at Northwestern University, University of California at Irvine and University of California at Los Angeles. Since 2021, he is an Associate Professor (tenure-track, civil-servant) of Computer Science with the Department of Computer Architecture and System Engineering, UCM. His main research interests include Design Automation, Next Generation Arithmetic and Quantum Computing. Dr. del Barrio has been the PI of the PARNASO project, funded by the Leonardo Grants program by Fundación BBVA, and currently, he is the PI of the ASIMOV project, funded by the Spanish MICINN, which includes a work package to research on the deployment of posits on RISC-V cores. Since 2019 he is an IEEE Senior Member and since December 2020, he is an ACM Senior Member, too.
\end{IEEEbiography}


\begin{IEEEbiography}[{\includegraphics[width=1in,height=1.25in,clip,keepaspectratio]{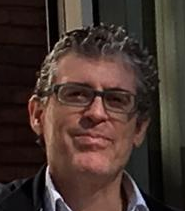}}]{Manuel Prieto-Matias}
Manuel Prieto Matias obtained a Ph.D. degree from Complutense University of Madrid (UCM) in 2000.  Since 2002, he has been a Professor at the Department of Computer Architecture at UCM, being a Full Professor since 2019. His research interests include high-performance computing, non-volatile memory technologies, accelerators, and code generation and optimization. His current focus is on effectively managing resources on emerging computing platforms, emphasizing the interaction between the system software and the underlying architecture. Manuel has co-authored over 100 scientific publications in journals and conferences in parallel computing and computer architecture. He is a member of the ACM.
\end{IEEEbiography}




\end{document}